\documentclass[twocolumn,aps,prb,showpacs,floatfix]{revtex4}

\usepackage{graphicx}
\usepackage{amsfonts}

\newcommand{\nn}{\nonumber}

\newcommand{\tg}{\tilde{g}}
\newcommand{\la}{\langle}
\newcommand{\ra}{\rangle}
\newcommand{\bea}{\begin{eqnarray}}
\newcommand{\eea}{\end{eqnarray}}

\begin{document}
\title{Theory of Two-Dimensional Josephson Arrays in a 
Resonant Cavity}
\author{E. Almaas}
\email{Almaas.1@nd.edu}
\altaffiliation{Present address: Deptartment of Physics, 
University of Notre Dame, Notre Dame, Indiana 46556.}
\author{D. Stroud}
\email{stroud@mps.ohio-state.edu.}
\affiliation{Department of Physics, The Ohio State University, Columbus,
Ohio 43210}

\date{\today}
\begin{abstract}
We consider the dynamics of a two-dimensional array of underdamped
Josephson junctions placed in a single-mode resonant cavity.  Starting
from a well-defined model Hamiltonian, which includes the effects of
driving current and dissipative coupling to a heat bath, we write down
the Heisenberg equations of motion for the variables of the Josephson
junction and the cavity mode, extending our previous one-dimensional
model.  In the limit of large numbers of photons, these equations can
be expressed as coupled differential equations and can be solved
numerically.  The numerical results show many features similar to
experiment.  These include (i) self-induced resonant steps (SIRS's) at
voltages $V = n\hbar\Omega/(2e)$, where $\Omega$ is the cavity
frequency, and $n$ is generally an integer; (ii) a threshold number
$N_c$ of active rows of junctions above which the array is coherent;
and (iii) a time-averaged cavity energy which is quadratic in the
number of active junctions, when the array is above threshold.  Some
differences between the observed and calculated threshold behavior are
also observed in the simulations and discussed.  In two dimensions, we
find a conspicuous polarization effect: if the cavity mode is
polarized perpendicular to the direction of current injection in a
square array, it does not couple to the array and there is no power
radiated into the cavity.  We speculate that the perpendicular
polarization would couple to the array, in the presence of
magnetic-field-induced frustration.  Finally, when the array is biased
on a SIRS, then, for given junction parameters, the power radiated
into the array is found to vary as the square of the number of active
junctions, consistent with expectations for a coherent radiation.  For
a given step, a two-dimensional array radiates much more energy into
the cavity than does a one-dimensional array.
\end{abstract}

\pacs{05.45.Xt, 79.50.+r, 05.45.-a, 74.40.+k}

\maketitle

\section{Introduction}

The properties of arrays of Josephson junctions have been of great
interest for nearly twenty years.\cite{review} Such arrays are
excellent model systems in which to study such phenomena as phase
transitions and quantum coherence in two dimensions.  For example, if
only the Josephson coupling energy is considered, the Hamiltonian of a
two-dimensional (2D) array of Josephson junctions is formally
identical to that of a 2D XY model [see, e. g. Ref.\
\onlinecite{chaikin}].  Arrays sometimes appear to mimic behavior seen
in nominally homogeneous materials, such as high-$T_c$
superconductors, which often behave as if they are composed of
distinct superconducting regions linked together by Josephson
coupling.\cite{tinkham} Finally, the arrays are of potentially
practical interest: they may be useful, for example, as sources of
coherent microwave radiation if the individual junctions can be caused
to oscillate in phase in a stable manner.

Recently, our ability to achieve this kind of stable oscillation, and
coherent microwave radiation, was significantly advanced by a series
of experiments by Barbara and
collaborators.\cite{barbara,barbara02,asc2000,vasilic02,vasilic} These
workers placed two-dimensional underdamped Josephson arrays in a
geometry which allowed them to be coupled to a resonant microwave
cavity.  The presence of the cavity caused the junctions to couple
together far more efficiently than in its absence.  As a result, the
power radiated into the cavity has been found to be as much as 30\% of
the d.\ c.\ power injected into the array, far higher than the
efficiency achieved in previous experiments.  Even more surprising,
this efficiency is achieved in {\em underdamped} arrays, which
according to conventional wisdom should be especially difficult to
synchronize, since each such junction exhibits bistability and
hysteresis as a function of the external control parameters.  These
experiments have stimulated many theoretical attempts to explain
them.\cite{filatrella,filatrella2,almaas,almaas02_2}

In our previous work, we have presented a simple {\em one-dimensional}
(1D) model which seems to account for many features of the observed
cavity-induced coherence.\cite{almaas,almaas02_2} Despite the
geometrical differences, the 1D model does a surprisingly good job of
capturing the physics of the experiments.  However, a truly realistic
test requires that the model be extended to a geometry closer to the
experimental one.  In this paper, therefore, we present the necessary
extension to 2D.  Our results give significant insight into why the 1D
model works so well.  In addition, they provide some clues about how
one might understand experimental features which are still unexplained
in either the 1D or the 2D models.

The remainder of this paper is organized as follows.  In the next
section, we describe our model Hamiltonian for a 2D current-driven,
underdamped Josephson junction array in a resonant cavity which
supports a single mode.  This Hamiltonian is a straightforward
extension of that used in our previous work to describe 1D arrays.  In
Section III, using this Hamiltonian, we write out the Heisenberg
equations of motion for the junction variables and for the photon
creation and annihilation operators for the cavity mode.  We
incorporate resistive dissipation in the junctions in a standard way,
by coupling the gauge-invariant phase differences across each junction
to its own set of harmonic oscillator variables whose spectral density
is chosen to produce Ohmic dissipation.  In the limit of large numbers
of photons, we obtain classical equations of motion for the variables.
In Section IV, we present the numerical solutions of this model with
an emphasis on features special to 2D, and we also give a comparison
between the 2D and previous 1D results.  A concluding discussion and
comparison with experiment follows in Section V.

\section{Model Hamiltonian}

\begin{figure}[t]
\centerline{\includegraphics[height=3.2cm]{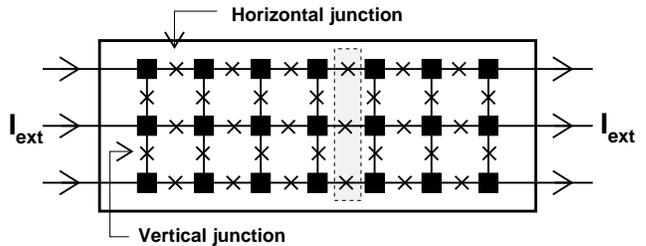}}
\caption{Sketch of the array geometry considered in our model.  There
are $(M\times N)$ superconducting islands (black squares), making
$\protect{[}(M-1)\times N+(N-1)\times M \protect{]}$ Josephson
junctions (crosses).  An external current $I^{ext}$ is injected into
each junction at one end of the array and extracted from each junction
at the other end.  The array is placed in an electromagnetic cavity
which supports a single resonant photon mode of frequency $\Omega$.
We have indicated by dashes a group of junctions which we denote a
``row.''  Such a row is perpendicular to the current bias and is
comprised of horizontal junctions.}
\label{fig:2Dgeometry}
\end{figure}

We will consider a 2D array of $N \times M$ superconducting grains
placed in a resonant cavity, which we assume supports only a single
photon mode of frequency $\Omega$.  The array is thus made up of
$(N-1)(M-1)$ square plaquettes.  There are a total of $N_x \times N_y$
horizontal junctions, where $N_x = N-1$ and $N_y = M$.  A current $I$
is fed into each of the $M$ grains on the left edge of the array, and
extracted from each of the $M$ grains on the right edge.  Thus, the
current is inject in the $x$ direction, with no external current
injected in the $y$ direction.  A sketch of this geometry is shown in
Fig. \ref{fig:2Dgeometry}.  We also introduce the terminology that a
``row'' of junctions, in this configuration, refers to a group of
$N_y$ junctions, all with left-hand end having the same $x$
coordinate, and all being parallel to the bias current.  One such row
is indicated by the dashed lines in Fig.\ \ref{fig:2Dgeometry}.

In contrast to our previous work,\cite{almaas02_2} we will write the
equations of motion for the grain variables (phases and charges)
rather than junction variables, since in 2D, the junction variables
cannot be treated as all independent (there are twice as many
junctions as grains).

We express our Hamiltonian in a form analogous to that of Ref.
\onlinecite{almaas02_2}:
\begin{equation}
H = H_{photon} + H_J + H_C + H_{curr} + H_{diss}. \label{eq2d:ham}
\end{equation}
Here $H_{photon}$ is the energy of the cavity mode, expressed as
\begin{equation}
H_{photon} = \hbar\Omega\left(a^\dag a + \frac{1}{2}\right),
\end{equation}
where $a^\dag$ and $a$ as the usual photon creation and annihilation
operators.  $H_J$ is the Josephson coupling energy, and is assumed to
take the form
\begin{equation}
H_J = -\sum_{\langle ij \rangle} E^J_{ij}\cos(\gamma_{ij}),
\end{equation}
where $E^J_{ij}$ is the Josephson energy of the (ij)$^{th}$ junction,
and $\gamma_{ij}$ is the gauge-invariant phase difference across that
junction (defined more precisely below).  $E^J_{ij}$ is related to
$I^c_{ij}$, the critical current of the (ij)$^{th}$ junction, by
$E^J_{ij} = \hbar I_{ij}^c /q$, where $q = 2|e|$ is the Cooper pair
charge.  $H_C$ is the capacitive energy of the array, which we write
in a rather general form as
\begin{equation}
H_C = \frac{1}{2}\sum_{ij}q^2(C^{-1})_{ij}n_in_j,
\end{equation}
where $C^{-1}$ is the inverse capacitance matrix, $n_i$ is the number
of Cooper pairs on the $i^{th}$ grain, and $q = 2e$ is the charge of a
Cooper pair (we take $e > 0$). Note that in
Ref. \onlinecite{almaas02_2}, the variable $n_i$ was used to denote
the {\em difference} between the numbers of Cooper pairs on the two
grains comprising junction $i$.

As in 1D, the gauge-invariant phase difference, $\gamma_{ij}$, is the
term which leads to coupling between the Josephson junctions and the
cavity.  We write it as
\begin{equation}
\gamma_{ij} = \phi_i - \phi_j - [(2\pi)/\Phi_0] \int_{ij} {\bf A}
	\cdot {\bf ds} \equiv~ \phi_i-\phi_j - A_{ij},
	\label{eq2d:gauge}
\end{equation} 
where $\phi_i$ is the gauge-dependent phase of the superconducting
order parameter on grain $i$, and ${\bf A}$ is the vector potential,
which (in Gaussian units) takes the form\cite{slater,yariv}
\begin{equation}
{\bf A}({\bf x},t) = \sqrt{(h c^2) / (\Omega)} \left(a(t) +
       a^\dag(t)\right){\bf E}({\bf x}), \label{eq:A}
\end{equation}
where ${\bf E}({\bf x})$ is a vector proportional to the local
electric field of the mode, normalized such that $\int_Vd^3x|{\bf
E}({\bf x})|^2 = 1$, and $V$ is the cavity volume.  The line integral
is taken across the (ij)$^{th}$ junction.

Given this representation for ${\bf A}$, the phase factor $A_{ij}$ can
be written
\begin{equation}
A_{ij} = g_{ij} (a + a^\dag),
\end{equation}
where $g_{ij}$ takes the form
\begin{equation}
g_{ij} = \sqrt{\frac{\hbar c^2}{\Omega} \frac{(2\pi)^{3}} {V\Phi_0^2}}
\int_{ij}{\bf E}\cdot{\bf ds}
\end{equation}
Clearly, $g_{ij}$ is an effective coupling constant describing the
interaction between the (ij)$^{th}$ junction and the cavity.

We include a driving current and dissipation in a manner similar to
that of Ref. \onlinecite{almaas02_2}.  The driving current is
included via a ``washboard potential,'' $H_{curr}$, of the form
\begin{equation}
H_{curr} = -\frac{\hbar I^{ext}}{q}\sum_{\langle ij\rangle \| \bf{ \hat{x}}}
	\gamma_{ij}, \label{eq2d:wash}
\end{equation}
where $I$ is the driving current injected in the $x$ direction into
{\em each} grain on the left edge (and extracted from the right edge),
and the sum runs over only those bonds in the $x$ direction (each such
bond is counted {\em once}).  To introduce dissipation, each
gauge-invariant phase difference, $\gamma_{ij}$, is coupled to a
separate collection of harmonic oscillators with a suitable spectral
density.\cite{ambeg,caldeira81,leggett,chakra} Thus, the dissipative
term in the Hamiltonian is
\begin{equation}
H_{diss} = \sum_{\langle ij \rangle} H_{ij}^{diss},
\end{equation}
where the sum runs over distinct bonds $\langle ij \rangle$, and
\[ H_{ij}^{diss} ~=~ \sum_{\alpha}\Bigg[f_{\alpha,ij}~ \gamma_{ij}~ u_{\alpha,ij} +
	\frac{(p_{\alpha,ij})^2}{2 m_{\alpha,ij}} +\]
\begin{equation}
\frac{1}{2}  m_{\alpha,ij}~ (\omega_{\alpha,ij})^2 ~(u_{\alpha,ij})^2
    + \frac{(f_{\alpha,ij})^2}{ 2~ m_{\alpha,ij}
    ~(\omega_{\alpha,ij})^2}~ (\gamma_{ij})^2 \Bigg].
    \label{eq2d:diss}
\end{equation}

The variables $u_{\alpha,ij}$ and $p_{\alpha,ij}$, describing the
$\alpha^{th}$ oscillator in the (ij)$^{th}$ junction, are canonically
conjugate, and $m_{\alpha,ij}$ and $\omega_{\alpha,ij}$ are the mass
and frequency of that oscillator.  By choosing the spectral density,
$J_{ij}(\omega)$, to be linear in $|\omega|$, we assure that the
dissipation in the junction is ohmic.\cite{caldeira81,leggett} We
write such a linear spectral density as
\begin{equation}
J_{ij}(\omega) = \frac{\hbar}{2\pi}~ \alpha_{ij}~ |\omega|~ \Theta (\omega_c -
              \omega ),
\end{equation}
where $\omega_c$ is a high-frequency cutoff (at which the assumption
of ohmic dissipation begins to break down), $\Theta(\omega_c -
\omega)$ is the usual step function, and $\alpha_{ij}$ is a
dimensionless constant. We write it as $\alpha_{ij} = R_0/R_{ij}$,
where $R_0 = h/(4e^2)$ and $R_{ij}$ is a constant with dimensions of
resistance (which proves to be the effective shunt resistance of the
junction, as discussed below).

\section{Equations of motion}

To obtain equations of motion, it is convenient to introduce the
operators $a = a_R + ia_I$ and $a^\dag = a_R - ia_I$.  These have the
commutation relation $[a_R, a_I] = i/2$, which follows $[a, a^\dag] =
1$.  In terms of these variables,
\begin{equation}
H_{photon} = \hbar\Omega(a_R^2 + a_I^2), \label{eq2d:ephot}
\end{equation}
and $\gamma_{ij}$ takes the form
\begin{equation}
\gamma_{ij} = \phi_i - \phi_j - 2 g_{ij} a_R.
\label{eq2d:gamma}
\end{equation}

The time-dependence of the various operators appearing in the
Hamiltonian (\ref{eq2d:ham}) is now obtained from the Heisenberg
equations of motion.  These are readily derived from the commutation
relations for the various operators in the Hamiltonian
(\ref{eq2d:ham}).  Besides the relations already given, the only
non-zero commutators are
\begin{eqnarray}
\protect{[}n_j, \phi_k\protect{]} &=& -i \delta_{jk};\\
\protect{[}p_{\alpha,ij}, u_{\beta,k\ell}\protect{]} &=& 
    -i\hbar~ \delta_{\alpha,\beta}~ \delta_{ij,k\ell},
\end{eqnarray}
\vspace*{.1cm}

\noindent
where the last delta function vanishes unless $(ij)$ and $(k\ell)$
refer to the {\em same} junction.

Using all these relations, we find, after a little algebra, the
following equations of motion for the operators $\phi_i$, $n_i$,
$a_R$, and $a_I$:
\begin{widetext}
\begin{eqnarray}
\dot{\phi}_i &=& \frac{q^2}{\hbar}\sum_j(C^{-1})_{ij}n_j,\label{eq2d:eom1} \\
\dot{n}_i  &=& -\frac{1}{\hbar} \sum_l E^J_{il} \sin(\phi_i- \phi_l- 2
      g_{il} a_R) +\frac{I_{i}^{ext}}{q} - \frac{1}{\hbar}\sum_{l}
      \sum_\alpha \left[u_{\alpha,il}f_{\alpha,il} +
      \frac{(f_{\alpha,il})^2}{m_{\alpha,il}(\omega_{\alpha,il})^2}
      (\phi_i-\phi_l -2 g_{il} a_R)\right],\label{eq2d:eom2} \\
\dot{a}_R &=& \Omega~ a_I,\label{eq2d:eom3} \\
\dot{a}_I &=& -\Omega~ a_R + \sum_{\langle ij \rangle} g_{ij}
      \frac{E^J_{ij}}{\hbar}~ \sin(\phi_i-\phi_j - 2 g_{ij}a_R) -
      \frac{I^{ext}}{q} \sum_{\langle ij \rangle \|{\bf
      \hat{x}}}g_{ij} \nn \\ && +\sum_{\langle ij \rangle} \frac{
      g_{ij}}{\hbar}~ \sum_{\alpha} \left(f_{\alpha,ij}~ u_{\alpha,ij}
      + \frac{(f_{\alpha,ij})^2} {m_{\alpha,ij}
      \omega_{\alpha,ij}^2}~ (\phi_i-\phi_j-2 g_{ij} a_R)
      \right).\label{eq2d:eom4}
\end{eqnarray}
\end{widetext}
Here, the index $l$ ranges over the nearest-neighbor grains of $i$.
In writing these equations, we have assumed that the only external
currents $I_{i}^{ext}$ are those along the left and right edges of the
array, where they are $\pm I^{ext}$ [cf.\ Fig.\ \ref{fig:2Dgeometry}].
Eqs.\ (\ref{eq2d:eom1})-(\ref{eq2d:eom4}) are equations of motion for
the {\em operators} $a_R$, $a_I$, $n_j$, and $\phi_j$ (or
$\gamma_j$). In order to make these equations amenable to computation,
we will later regard these operators as $c$-numbers, as we did earlier
in 1D.\cite{almaas02_2} This approximation is expected to be
reasonable when there are many photons in the cavity.\cite{almaas02_2}

The equations of motion for the harmonic oscillator variables can also
be written out explicitly.  However, since we have no direct interest
in these variables, we instead eliminate them in order to incorporate
a dissipative term directly into the equations of motion for the other
variables.  Such a replacement is possible provided that the spectral
density of each junction is linear in frequency, as noted above.  In
that case,\cite{ambeg,caldeira81,leggett,chakra,almaas02_2} the
oscillator variables can be integrated out.  The effect of carrying
out this procedure is that one should make the replacement
\begin{equation}
\sum_{\alpha} \left( f_{\alpha,ij}~ u_{\alpha,ij} +  \frac{(f_{\alpha,ij})^2}
	{m_{\alpha,ij} \omega_{\alpha,ij}^2}~ \gamma_{ij} \right) \rightarrow
       \frac{\hbar}{2\pi}\frac{R_0}{R_{ij}}~ \dot{\gamma}_{ij} 
       \label{eq2d:subst}
\end{equation}
wherever this sum appears in the equations of motion.  Making the
replacement (\ref{eq2d:subst}) in Eqs.\ (\ref{eq2d:eom2}) and
(\ref{eq2d:eom4}), and simplifying, we obtain the equations of motion
for $n_j$ and $a_I$ with damping:
\begin{widetext}
\begin{eqnarray}
\dot{n}_i &=& -\sum_j\frac{E^J_{ij}}{\hbar} \sin(\gamma_{ij}) +
        \frac{I^{ext}_{i}}{q} -\sum_j\frac{1}{2\pi}\frac{R_0}{R_{ij}}
        \dot{\gamma}_{ij} \label{eq2d:eom2a} \\
\dot{a}_I &=& -\Omega~ a_R + \sum_{\langle ij \rangle} g_{ij}
        \frac{E_{ij}}{\hbar}~ \sin(\gamma_{ij}) - \frac{I^{ext}}{q}
        \sum_{\langle ij \rangle \| {\bf \hat{x}}} g_{ij} +
        \sum_{\langle ij \rangle} g_{ij}~ \frac{R_0}{2\pi R_{ij}}
        \dot{\gamma}_{ij}.
\label{eq2d:eom4a}
\end{eqnarray}
\end{widetext}
Once, again, the index $j$ is summed only over the nearest-neighbor
grains of $i$.  Equations (\ref{eq2d:eom1}), (\ref{eq2d:eom3}),
(\ref{eq2d:eom2a}), and (\ref{eq2d:eom4a}) form a closed set of
equations which can be solved for the time-dependent functions
$\gamma_i$, $n_i$, $a_R$ and $a_I$, given the external current and the
other parameters of the problem.

It is now convenient to express these equations of motion in terms of
suitable scaled variables.  We therefore introduce a dimensionless
time $\tau = t q R I^c /\hbar = \omega_{\tau} t$, where $R$ and $I^c$
are suitable averages over $R_{ij}$ and $I^c_{ij}$. We also define the
other scaled variables
\begin{eqnarray}
\tilde{R}_{ij} &=& \frac{R_{ij}}{R}, \\
\tilde{\Omega} &=& \frac{\Omega}{\omega}, \\
\tilde{I}      &=& \frac{I}{I^c}, \\
\tilde{V}_i    &=& \frac{V_i}{R I^c}, \label{eq2d:pot}\\
\tilde{a}_{R,I}&=& \sqrt{2 \pi \frac{R}{R_0}} a_{R,I}, \\
\tilde{g}_{ij} &=& \sqrt{\frac{R_0}{2\pi R}} g_{ij},\\
\tilde{C}_{ij} &=& \omega_{\tau} R C_{ij}.
\end{eqnarray}
The last equation involves the capacitance matrix $C_{ij}$.  We assume
that this takes the form \cite{fazio,kim}
\begin{eqnarray}
C_{ij} &=& \left(C_d + z_i~C_c \right) \delta_{ij} \nonumber \\ &&- C_c
    \left(\delta_{i,j+{\bf \hat{x}}} + \delta_{i,j-{\bf \hat{x}}} +
    \delta_{i,j+{\bf \hat{y}}} + \delta_{i,j-{\bf \hat{y}}}\right),
\end{eqnarray}
i.\ e., that there is a nonvanishing capacitance only between
neighboring grains and between a grain and ground. Here $z_i (=4)$ is
the number of nearest neighbors of grain $i$, $C_d$ and $C_c$ are
respectively the diagonal (self) and nearest-neighbor capacitances,
and ${\bf \hat{x}}$ and ${\bf \hat{y}}$ are unit vectors in the $x$
and $y$ directions.  The corresponding Stewart-McCumber parameters are
$\beta_c = \omega_{\tau} R C_c$ and $\beta_d = \omega_\tau R C_d$.

In Eq.\ (\ref{eq2d:pot}), we introduced the potential $V_i$ on site
$i$, which is expressed through the number variables $n_j$ as
\begin{equation}
V_i ~=~ q \sum_j (C^{-1})_{ij} n_j.
\end{equation}
The integral of the electric field across junction $(ij)$ is written
in terms of the $V_i$'s as
\begin{equation}
V_{ij} ~=~ V_i - V_j - 2 \tg_{ij} \Omega a_I.
\end{equation}

Carrying out these variable changes, we find, after some algebra, that
the equations of motion can be expressed in the following
dimensionless form:

\begin{widetext}
\begin{eqnarray}
\frac{d}{d\tau}~\phi_i      &=& \tilde{V}_i \label{eq:2dsolve1},\\
\frac{d}{d\tau}~\tilde{V}_i &=& \sum_j (\tilde{C}^{-1})_{ij} \Bigg[
    \tilde{I}_j^{ext} - \sum_l \Bigg( \tilde{I}_{j l}^c \sin (\phi_j -
    \phi_l - 2 \tilde{g}_{j l} a_R ) + \frac{1}{\tilde{R}_{j l}}
    (\tilde{V}_i - \tilde{V}_l - 2 \tilde{g}_{i l} \tilde{\Omega}
    \tilde{a}_I)\Bigg) \Bigg], \\
\frac{d}{d\tau}~\tilde{a}_R &=& \tilde{\Omega} \tilde{a}_I, \\
\frac{d}{d\tau}~\tilde{a}_I &=&-\tilde{\Omega} \tilde{a}_R +
    \sum_{\langle ij \rangle} \tilde{g}_{ij} \Bigg[ \tilde{I}_{i j}^c
    \sin (\phi_i\! -\!  \phi_j \!-\! 2 \tilde{g}_{i j} \tilde{a}_R) +
    \frac{1}{\tilde{R}_{i j}} (\tilde{V}_i - \tilde{V}_j - 2
    \tilde{\Omega} \tilde{g}_{i j} \tilde{a}_I)\Bigg] -
    \tilde{I}^{ext} \!\!\! \sum_{\langle ij \rangle \| {\bf
    \hat{x}}}\!\! \tilde{g}_{i j} \label{eq:2dsolve2}.
\end{eqnarray}
\end{widetext}
These equations are readily generalized to treat external currents
with non-zero components in both the $x$ and the $y$ directions, and
to geometries other than lattices with square primitive cells.

\section{Numerical results}

We solve Eqs.\ (\ref{eq:2dsolve1}) -- (\ref{eq:2dsolve2}) numerically,
by implementing the adaptive Bulrisch-Stoer method,\cite{numrec} as
described further in Ref.\ \onlinecite{almaas02_2}.  For simplicity,
we assume that the coupling constants $\tg_{ij}$ have only two
possible values, $\tg_x$ and $\tg_y$, corresponding to junctions in
the $x$ and $y$ direction respectively.\cite{note1} This assumption
should be reasonable if two conditions are satisfied: (i) there is not
much disorder in the characteristics of the individual junctions; and
(ii) the wavelength of the resonant mode is large compared to the
array dimensions.  Although assumption (ii) is not obviously satisfied
for the experimental arrays, the model may still be reasonable in
certain array/cavity geometries, as discussed further below.

In the presence of a vector potential, it is customary to define a
{\em frustration} $f_\mu(\tau)$ for the $\mu^{th}$ plaquette by the
relation\cite{teitel}
\begin{eqnarray}
f_\mu(\tau) &=& \frac{1}{2\pi} \! \sum_{plaquette} \!\!\!\! A_{ij} ~=~ 
\frac{\tilde{a}_R(\tau)}{\pi} \!\! \sum_{plaquette}\!\!\!\! \tg_{ij},
\end{eqnarray}
where the sum runs over bonds in the $\mu^{th}$ plaquette.  For a
general position-dependent $\tg_{ij}$, $f_\mu(\tau) \neq 0$, but if
$\tg_x$ and $\tg_y$ are both position-independent, then $f_\mu(\tau) =
0$.  The possibility of having distinct coupling constants $\tg_x$ and
$\tg_y$ along $x$ and $y$ bonds arises from differences in possible
polarizations of the resonant mode, and leads to effects which cannot
be captured in a 1D model, as discussed below.

Before discussing our numerical results, we briefly summarize one
well-known feature of underdamped Josephson arrays in the {\em
absence} of coupling to a resonant cavity.  At certain applied
currents, the individual junctions in such an array are bistable -
that is, they can be placed in an ``active'' (resistive) or an
``inactive'' (superconducting) state, by a careful choice of initial
conditions.  For an applied current in the $x$ direction, when a
single horizontal junction is chosen to be in the active state, it is
found that all the other horizontal junctions in the same ``row''
(cf.\ Fig.\ \ref{fig:2Dgeometry}) also go active, provided that there
is at least a little disorder in the junction critical currents [cf.,
e.\ g., Refs.\ \onlinecite{wenbin1} and \onlinecite{wenbin92}].  In
our simulations for 2D arrays coupled to a resonant cavity, we observe
this same phenomenon, as discussed below.

\subsection{Horizontal coupling}

\begin{figure}[t]
\centerline{\includegraphics[height=7cm]{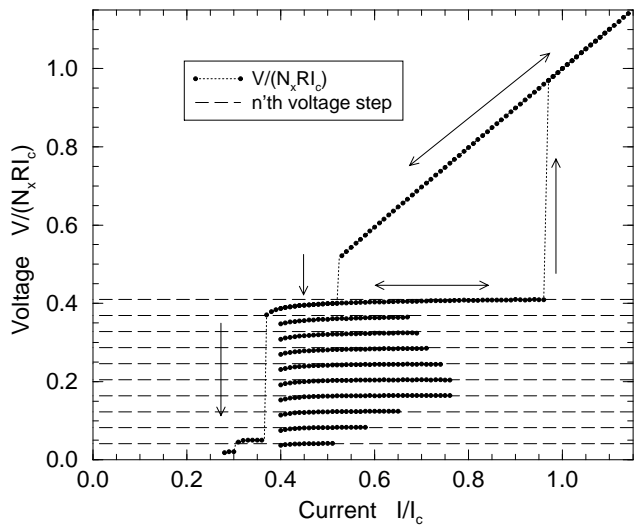}}
\caption{Calculated current-voltage characteristic for a $10\times4$
array with cavity frequency $\tilde{\Omega} = 0.41$, capacitance
parameters $\beta_c = 20$ and $\beta_d = 0.05$, disorder parameter
$\Delta = 0.05$ and junction-cavity coupling in the horizontal
direction $\tg_{x} = 0.012$. The horizontal dashed lines show voltages
at which the various SIRS's are expected. These correspond to
different numbers of rows of horizontal junctions in the active
state. Arrows denote that the given IV was taken in the direction of
increasing or decreasing current.}
\label{fig:10x4_IV}
\end{figure}

We first consider the case $\tg_x \neq 0$, $\tg_y = 0$, with driving
current parallel to the $x$ axis.  In Fig.\ \ref{fig:10x4_IV}, we show
a series of current-voltage (IV) characteristics for this case.  We
consider an array of $10\times 4$ grains, with capacitances $\beta_c =
20$ and $\beta_d = 0.05$, $\tg_x = 0.012$, and $\tilde{\Omega} =
0.41$. The critical current through the (ij)$^{th}$ junction is
$\tilde{I}_{i j}^c = 1 + \Delta_{ij}$ where the disorder $\Delta_{ij}$
is randomly selected with uniform probability from
$[-\Delta,\Delta]$. In this plot, $\Delta = 0.05$. The product
$\tilde{I}_{i j}^c \tilde{R}_{ij}$ is assumed to be the same for all
junctions, in accordance with the Ambegaokar-Baratoff
expression.\cite{ambegaokar} In addition, $\beta_d$ and $\beta_c$ are
assumed to be the same for all junctions.  The calculated IV's are
shown as a series of points.  The directions of the arrows indicate
whether the curves were obtained under increasing or decreasing
current drive, or both.  The horizontal dashed curves correspond to
voltages where {\em self-induced resonant steps} (SIRS's) are
expected, namely $\langle V \rangle_{\tau}/(NRI_c) =\tilde{\Omega}$ in
our units, where $\langle V \rangle_{\tau}$ denotes the time-averaged
voltage. The dotted lines are guides to the eye.  Each nearly
horizontal series of points denotes a calculated IV characteristic for
a {\em different} number of active rows $N_a$, and represents $N_a
\times N_y$ (horizontal) junctions sitting on the first integer ($n =
1$) SIRS.  The calculated voltages for the various $N_a$'s agree well
with the expected values given by the dashed horizontal lines.  The
long straight diagonal line segment, which is common to all the
different $N_a$'s, represents the ohmic part of the IV characteristic
with all rows active. For the sake of clarity, we have chosen not to
plot the corresponding segments for other choices of $N_a < 10$.
Besides the integer SIRS's we find that for this 2D array, it is
possible to bias individual active rows on either the $n = 1/2$ or the
$n = 2$ SIRS.  (A small segment of an $n = 1/2$ case is visible in the
lower left of the figure.)  Similar behavior is found in the case of
Shapiro steps produced by a combined d.\ c. and a.\ c. current in a
conventional underdamped Josephson junction (see, e.\ g.\ Ref.\
\onlinecite{waldram}).

Although the full hysteresis loop is shown in Fig.\ \ref{fig:10x4_IV}
only for $N_a = 10$ active rows, the IV curves for other values of
$N_a$ are also hysteretic.  Specifically (as also found previously in
the 1D case), whenever $N_a > 4$, the number of active rows increases
when the SIRS's become unstable.  That is, if the current is increased
so that a given SIRS becomes unstable, the IV characteristic jumps up
onto a higher SIRS, and also some of the individual rows jump onto the
$n = 2$ SIRS.  The IV curve only jumps onto the ohmic branch if $I/I_c
> 1$.  By contrast, if the applied current is changed so that the
SIRS's become unstable for $N_a \leq 4$, the number of active rows
remains unchanged and the IV curve immediately becomes ohmic. In this
regime, if $I$ is increased so that $I/I_c \sim 1$, all the remaining
horizontal junctions become active and the IV characteristic also
becomes ohmic. Another feature of these results worth noticing is
that the width of the SIRS plateaus is non-monotonic in $N_a$.  By
``width'' of an SIRS, we mean the range of driving currents for which
the SIRS is stable.

\begin{figure}[t]
\centerline{\includegraphics[height=7cm]{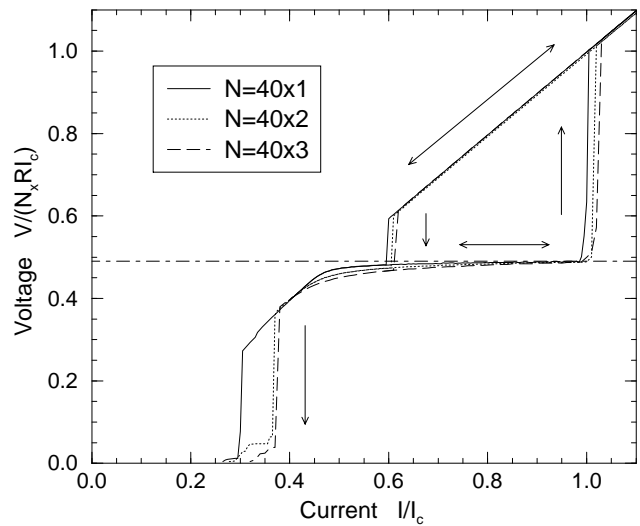}}
\caption{Calculated current-voltage characteristics for a $40\times 1$
(full line), a $40\times 2$ (dotted line) and a $40\times 3$
(long-dashed line) array, all with parameters $\tilde{g}_x = 0.015$,
$\tilde{\Omega} = 0.49$, $\beta_c = 20$, $\beta_d = 0.05$ and $\Delta
= 0.05$.  The horizontal dot-dashed line shows the expected position
of the SIRS.  Note that as the array width increases, the smallest
value of $\tilde{I}$ at which all the active junctions phase-lock on
the SIRS also increases, and IV characteristic on the SIRS has an
increasing bend.  Hence, increasing the array width at fixed $\tg_x$
has a effect similar to that of increasing $\tg_x$ at fixed width.
The arrows indicate the direction of the current sweep.}
\label{fig:40x_IV}
\end{figure}

Fig.\ \ref{fig:40x_IV} shows the IV characteristics for three
different arrays, each with all rows in the active state: (i) a
$40\times 1$ (full curve), (ii) a $40\times 2$ (dotted curve) and
(iii) a $40\times 3$ (long-dashed curve).  Each array has the
parameters $\tilde{g}_x = 0.015$, $\tilde{\Omega} = 0.49$, $\beta_c =
20$, $\beta_d = 0.05$ and $\Delta = 0.05$. Once again, the arrows
denote the directions of current sweep.  The horizontal dot-dashed
curve shows the expected position of the SIRS corresponding to $N_a =
40$ [$V/(N_xRI_c) = \tilde{\Omega}$].  The curves show that all three
arrays have qualitatively similar behavior.  First, if the array is
started from a random initial phase configuration, such that
$\tilde{I} \equiv I/I_c > 1+\Delta$, and $\tilde{I}$ is decreased,
then all the rows lock on to the $N_a = 40$ SIRS.  Secondly, if
$\tilde{I}$ is further decreased, the $N_a = 40$ active state
eventually becomes unstable and all the junctions go into their
superconducting states.  Finally, if $\tilde{I}$ is {\em increased}
starting from a state in which the array is on the $N_a = 40$ SIRS,
the SIRS remains stable until $\tilde{I}$ reaches the critical current
for the various rows, and the IV curve becomes ohmic.

The behavior shown in Fig.\ \ref{fig:40x_IV} with increasing array
width is very similar to that found previously in 1D arrays with
increasing {\em coupling strength}.  In other words, the key parameter
in understanding the curves of Fig.\ \ref{fig:40x_IV} is the product
$N_y\tg_x$.  For example, Fig.\ \ref{fig:40x_IV} shows that the effect
of increasing $N_y$ while keeping $\tg_x$ constant is to raise
slightly the maximum value of $\tilde{I}$ for which the active
junctions are still locked onto the $N_a = 40$ SIRS.  Furthermore,
that portion of the SIRS which corresponds to small $\tilde{I}$ is not
perfectly flat (i. e. not at the expected constant voltage
$V/(N_xRI_c)=\tilde{\Omega}$), but instead increases slightly with
increasing $\tilde{I}$ [cf. Fig.\ \ref{fig:40x_IV}].  The degree of
this non-flatness increases with increasing $N_y$.  Precisely
analogous effects are seen in calculations for 1D arrays with
increasing $\tg_x$.\cite{almaas02_2} This is another piece of evidence
that the key parameter is the product $N_y\tg_x$.

\begin{figure}[t]
\centerline{\includegraphics[height=7cm]{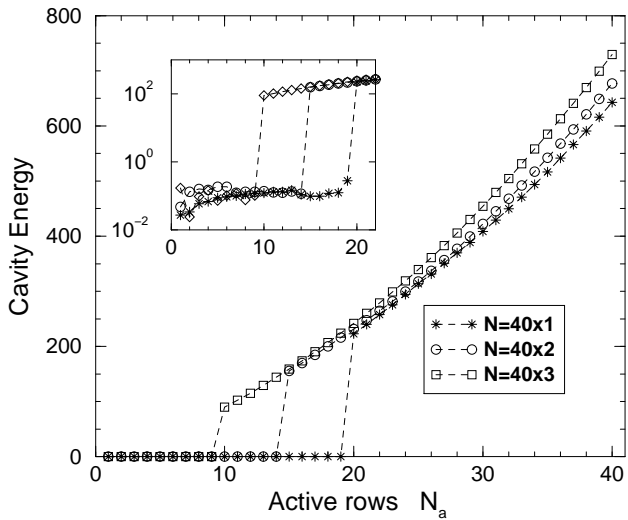}}
\caption{Time-averaged scaled energy $\tilde{E}$ in the resonant
cavity as function of active number of rows for a $40\times 1$
(asterisks), a $40\times 2$ (circles) and a $40\times 3$ (squares)
array with driving current $\tilde{I} = 0.58$. All the other
parameters are the same as those of Fig.\ \ref{fig:40x_IV}.  Inset: an
enlargement of the IV characteristics near the synchronization
threshold, on a logarithmic vertical scale.  Note that the threshold
number of active junctions for synchronization decreases with
increasing array width.}
\label{fig:40x_thresh}
\end{figure}

In Fig.\ \ref{fig:40x_thresh}, we plot the time-averaged energy
$\tilde{E}(N_a) = \la \tilde{a}_R^2 + \tilde{a}_I^2 \ra_{\tau}$ in the
cavity for three different arrays: $40\times 1$ (stars), $40\times 2$
(circles), and $40\times 3$ (squares).  In all cases, $\tilde{I} =
0.58$, and the other parameters are the same as those of
Fig. \ref{fig:40x_IV}. Below a threshold value of $N_a$, (which we
denote $N_c$ and which depends on $N_y$), the active rows are in the
McCumber state (not on the SIRS's).  In this case, $\tilde{E}(N_a)$ is
small and shows no obvious functional dependence on $N_a$ (see inset).
By contrast, above threshold, $\tilde{E}(N_a)$ is much larger and
increases as $N_a^2$.

Fig.\ \ref{fig:40x_thresh} shows that, when $N_y$ is increased at
fixed $\tg_x$, $N_c$ decreases.  Precisely this same trend is observed
when we increase $\tg_x$ while holding $N_y$ fixed (and was observed
in our previous 1D calculations with increasing $\tg_x$).  Thus, once
again, the relevant parameter in understanding the threshold behavior
appears to be $N_y \tg_x$.

As in 1D arrays, it is useful to introduce a {\em Kuramoto order
parameter} which describes the phase ordering.  For the 2D arrays, we
define a Kuramoto order parameter $\langle r_h\rangle_{\tau}$ for the
horizontal bonds by
\begin{equation}
\langle r_h \rangle_{\tau} = \frac{1}{N_a N_y}~\langle | \sum_{\la ij \ra
     \parallel \hat{\bf x}} e^{i\gamma_{ij}}| \rangle_{\tau},
     \label{eq:2Dkuramoto}
\end{equation}
where $N_a$ is the number of active rows, $N_y$ is the number of
horizontal junctions in a single row and the sum runs over all the
active, horizontal junctions.  [The analogous quantity $\langle
r_v\rangle_{\tau}$ for the vertical junctions is irrelevant when
$\tg_v = 0$, since in this case these junctions are inactive.]  For
the parameters shown in Fig.\ \ref{fig:40x_thresh}, we have found, as
in our previous 1D calculations, that $\langle r_h \rangle_{\tau} \sim
1$ for $N_a > N_c$ while $\langle r_h \rangle_{\tau} \ll 1$ for $N_a <
N_c$.  This behavior (which we do not show in a figure) reflects the
fact that, for the value of $\tilde{I}$ used in Fig.\
\ref{fig:40x_thresh}, none of the active junctions are on a SIRS when
$N_a < N_c$; hence, these junctions are not in phase with one another,
and the value of $\langle r_h \rangle_{\tau}$ reflects this lack of
coherence.

\begin{figure}[t]
\centerline{\includegraphics[height=6.8cm]{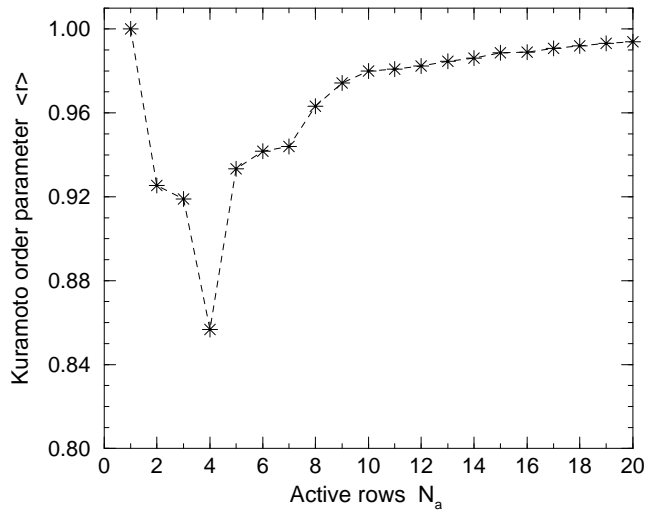}}
\caption{The time-averaged Kuramoto order parameter $\langle r_h
\rangle_\tau$ (defined in Eq. (\ref{eq:2Dkuramoto}) as a function of
the number of active rows on SIRS's for a $20\times 2$ Josephson array
with $\tilde{\Omega} = 0.49$, $\tg_x = 0.01$, $\beta_c = 20.$,
$\beta_d = 0.05$, $\Delta = 0.1$ and bias current $\tilde{I} = 0.53$.}
\label{fig:kuram_small}
\end{figure}

For certain array parameters, $\tilde{I}$ can be chosen so that {\em
all} the active junctions lie on SIRS's, however many active rows
$N_a$ there are.  [In Fig.\ \ref{fig:10x4_IV}, for example, $\tilde{I}
\sim 0.5$ would achieve this result.]  In such cases, even though all
the active junctions are oscillating with the same frequency, and
locked onto SIRS's, it is still possible to have $\langle r_h
\rangle_{\tau} < 1 $.  In this situation, the Kuramoto order parameter
$\langle r_{h,n}\rangle_{\tau} \sim 1$ for the {\em individual rows}.
This occurs because the {\em rows} are not perfectly phase-locked to
one other.  An example of such behavior is shown in Fig.\
\ref{fig:kuram_small}, for a $20\times 2$ junction array for several
numbers $N_a$ of active rows.  The other parameters are
$\tilde{\Omega} = 0.49$, $\tg_x = 0.01$, $\beta_c = 20.$, $\beta_d =
0.05$, $\Delta = 0.1$ and $\tilde{I} = 0.53$.  As the number of active
rows on the SIRS's increases, $\langle r_x \rangle_{\tau} \rightarrow
1$.  [Also, of course, $\langle r_x \rangle_{\tau} = 1$ for {\em one}
active row on a SIRS.]  Numerically, we find that it is easier in 2D
than in 1D to achieve a state with all active junctions biased on a
SIRS, but with $\la r_x \ra_\tau < 1$.  In all such cases, we can
easily cause $\la r_x \ra_\tau \rightarrow 1$ simply by increasing
$\tg_x$.

\begin{figure*}[t]
\centerline{\includegraphics[height=7cm]{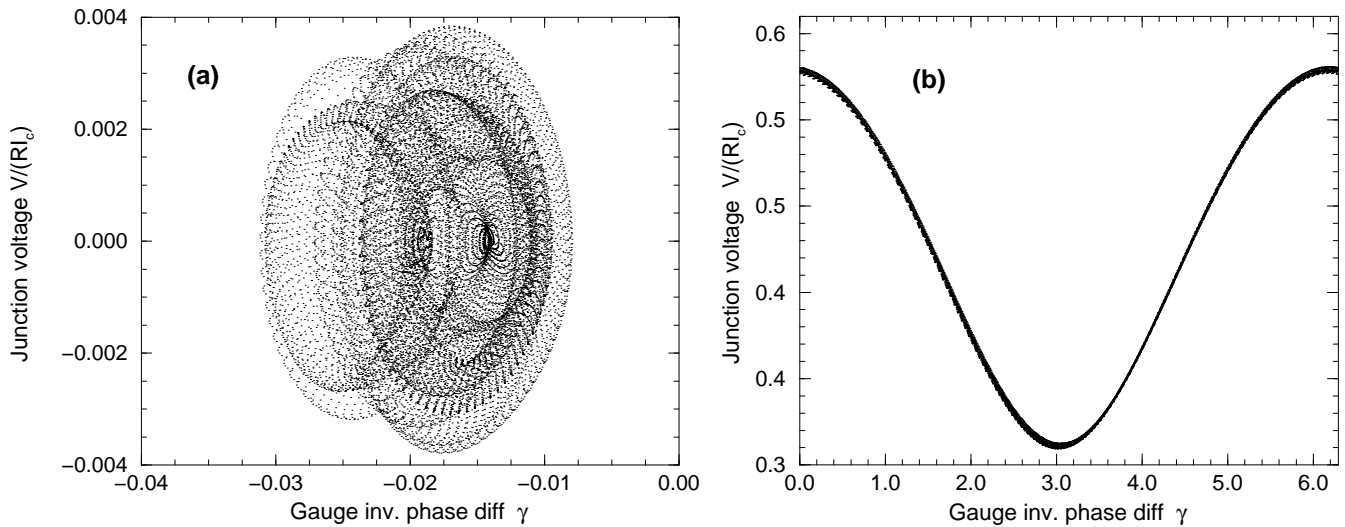}}
\caption{Phase plot of points $(\gamma, \dot{\gamma})$ where $\gamma$
is the gauge-invariant phase difference across a junction, shown for
(a) a vertical junction and (b) a horizontal junction, in an array in
which the cavity couples only to the horizontal junctions: $\tg_x =
0.$, $\tg_y = 0.5$.  The array size is $10\times 4$, and the other
parameters are $\tilde{\Omega} = 0.45$, $\Delta = 0.05$, $\beta_c =
20$, $\beta_d = 0.05$ and $\tilde{I} = 0.46$.  The vertical junction
in (a) displays aperiodic motion with very small amplitude,
corresponding to no time-averaged voltage drop across that junction,
while the horizontal junction in (b) has a phase difference which
varies periodically in time.}
\label{fig:2Dperiodicity}
\end{figure*}

The threshold shown in Fig.\ \ref{fig:40x_thresh} corresponds to a
transition from a state in which {\em none} of the active junctions
are on SIRS's to a state in which {\em all} are on SIRS's.  It is
possible to choose $\tilde{I}$ so as to have {\em any number} of
active rows $N_a$ on SIRS's.  In this case, the cavity energy
$\tilde{E}(N_a)$, in our model, is approximately quadratic in $N_a$,
with no obvious threshold behavior.  This feature of our results is
discussed further below.

\subsection{Vertical coupling}

We have also investigated the case of $\tg_x = 0$, $\tg_y \neq 0$, for
a wide range range of $\tg_y$ values.  For our geometry, we have not
been able to find {\em any} value for $\tg_y$ for which a SIRS
develops.  In essence, when the cavity couples {\em only} to the
vertical junctions, it is invisible in the IV characteristics.  This
behavior is easily understood.  In this geometry, with current applied
in the $x$ direction, both the time-averaged voltage and the
time-averaged current through the vertical junctions are very small.
Hence, too little power is dissipated in the vertical junctions to
induce a resonance with the cavity.

To illustrate this behavior, we show in Fig.\ \ref{fig:2Dperiodicity}
some representative phase plots of $(\gamma_{ij},\dot{\gamma}_{ij})$
for (a) a vertical junction and (b) a horizontal junction in a
$10\times 4$ array with $\tg_x = 0$, $\tg_y = 0.5$, $\beta_c = 20$,
$\beta_d = 0.05$, $\tilde{\Omega} = 0.45$, $\Delta = 0.05$ at bias
current $\tilde{I} = 0.46$ (close to a possible resonance with
cavity).  The phase plot for the vertical junction exhibits
small-amplitude aperiodic motion, while that of the horizontal
junction shows that this junction is in its active state and
undergoing periodic motion in phase space.  This lack of response by
the $y$ junctions to the cavity probably explains why the 1D
simulations describe the experiments so well.

It is no surprise that the cavity interacts only very weakly with the
vertical junctions.  From previous studies of both underdamped and
overdamped disordered Josephson arrays in a rectangular geometry (see,
e. g., Refs.\ \onlinecite{wenbin94}, and \onlinecite{wenbin92}), it is
known that when current is applied in the $x$ direction, the $y$
junctions remain superconducting, with $\la V\ra_{\tau} \approx 0$,
while the $x$ junctions comprising an active row are almost perfectly
synchronized, with $\la r_x \ra \approx 1$.

If there were an an external magnetic field {\em perpendicular} to the
array, we believe that SIRS's would be generated for $\tg_y \neq 0$,
even if $\tg_x = 0$.  In this case, the magnetic field would induce
frustration.  Specifically, since the sum of the gauge invariant phase
differences around a plaquette must be an integer multiple of $2\pi$,
the presence of magnetic-field-induced vortices piercing the
plaquettes would induce nonzero voltages across, and supercurrents in,
the $y$ junctions.  It would be of great interest if calculations were
carried out in such applied magnetic fields.

\subsection{Comparison with 1D Model}

We now compare our 2D results explicitly with those for 1D arrays.  In
our earlier 1D model, we found numerically\cite{almaas02_2} that the
threshold number of active junctions $N_c$ was inversely proportional
to the coupling constant $\tg$.  This behavior is reasonable because
the inhomogeneous term driving the cavity variable $\tilde{a}_R$ is
proportional to the product of $\tg$ and $N_a$.

Some of our numerical trends in the 2D case can be understood
similarly.  For example, the inhomogeneous term in Eq.\
(\ref{eq:2dsolve2}) is the last term on the right-hand side.  It is
proportional to the sum of the coupling constants $\tg_{ij}$ over {\em
all} the junctions parallel to $\tilde{I}$.  Thus, for $\tg_x \neq 0$, $\tg_y
= 0$, and for the same driving current $\tilde{I}$, we expect that an $N_x
\times N_y$ array with a coupling constant $\tg_x$ should behave like
an $N_x \times 1$ array with coupling constant $N_y\tg_x$.

\begin{figure}[t]
\centerline{\includegraphics[height=7cm]{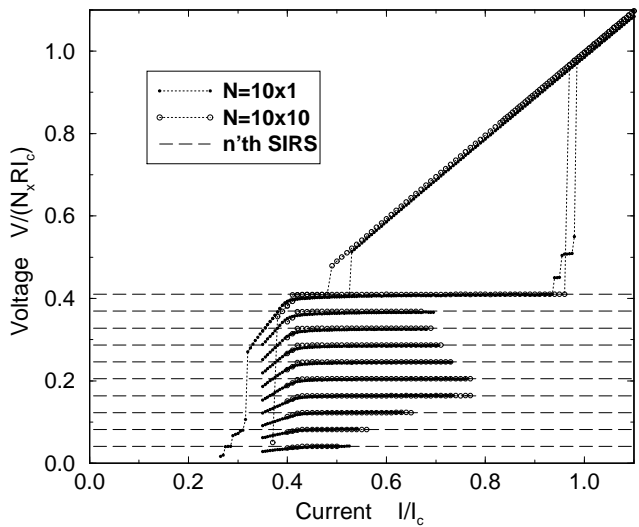}}
\caption{IV characteristics for a $10\times 1$ array ($\ast$) and a
$10\times 10$ array ($\circ$).  The $ 10 \times 1)$ array has
parameters $\tilde{g}_{x, 10\times 1} = 0.0259$, $\tilde{\Omega} =
0.41$, $\beta_c = 20$, $\beta_d = 0.05$ and $\Delta = 0.05$. The
expected position of the SIRS's are marked by horizontal dashed lines.
The $10\times 10$ array has $\tg_{x, 10\times 10}$ = 0.00259, and the
other parameters are the same as for the $10\times 1 $ array.  The IV
characteristics are shown for both increasing and decreasing current
drive, as discussed in the text.}
\label{fig:10x_IV}
\end{figure}

To check this hypothesis, we compare, in Fig.\ \ref{fig:10x_IV}, the
IV characteristics of a $10\times 1$ array having coupling constant
$\tg_{x;10\times 1} = 0.0259$ with those of a $10\times 10$ array with
coupling constant $\tg_{x;10\times 10} = 0.00259$.  The other
parameters are the same for the two arrays: $\tilde{\Omega} = 0.41$,
$\beta_c = 20$, $\beta_d = 0.05$ and $\Delta = 0.05$.  The expected
positions of the SIRS's [at $V/(NRI_c) = \tilde{\Omega}$] are
indicated by dashed horizontal lines.  Indeed, the two sets of IV
characteristics are very similar.  Even some of the subtle differences
can be understood in a simple way.  For example, the $10 \times 10$
IV's are slightly flatter than the $10 \times 1$ curves.  We believe
this extra flatness occurs because the individual junction couplings
in the $10 \times 1$ array are $10$ times larger than those in the $10
\times 10$ array.  From our previous 1D simulations, the IV's on the
steps become more and more rounded as $\tg_x$ increases, i. e., the
voltage on the lower portion of the SIRS is no longer independent of
$\tilde{I}$ [cf. Ref.\ \onlinecite{almaas02_2}].  Precisely this
behavior is seen in Fig.\ \ref{fig:10x_IV}.

Another subtle difference between the 1D and 2D curves of Fig.\
\ref{fig:10x_IV} is the values of the so-called ``retrapping current''
in the two sets of curves (i. e. the current values below which the
McCumber curve becomes unstable).  We believe that this difference can
be understood in terms of the effects of disorder in the junction
critical currents in 1D and 2D.  Specifically, for a given value of
$\Delta$, the 2D arrays are effectively less disordered than the 1D
arrays, since the {\em average} critical current for a single row has
a smaller rms spread than the critical current of a single junction in
a 1D array.

An important similarity between the two sets of curves is that, in
both the $10\times 10$ and the $10\times 1$ arrays, the width of the
SIRS's varies similarly (and non-monotonically) with the number of
active rows.  This behavior distinguishes our predictions from some
other models,\cite{filatrella,filatrella2} in which the cavity is
modeled as an RLC oscillator connected in parallel to the entire
array, and which predicts a monotonic dependence of SIRS width on
$N_a$. \cite{barbara02}

\begin{figure}[t]
\centerline{\includegraphics[height=7cm]{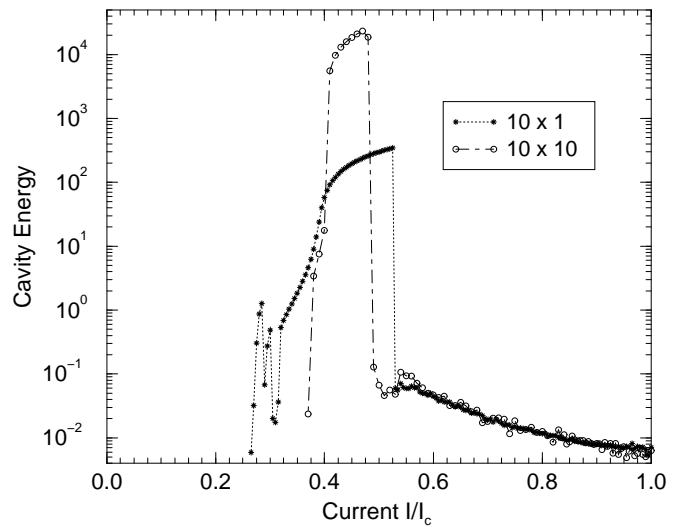}}
\caption{Time-averaged reduced cavity energy $\tilde{E}$, for a
$10\times 1$ array and a $10\times 10$ array for the same choice of
array parameters as in Fig.\ \ref{fig:10x_IV}.  The calculations are
carried out on the decreasing current branch with all rows active.
Note that $\tg_x$ for the $10\times 10$ array is 10 times smaller than
that of the $10\times 1$ array.}
\label{fig:10x_cav}
\end{figure}

In Fig.\ \ref{fig:10x_cav} we plot the reduced time-averaged cavity
energy $\tilde{E} = \la a_R^2 + a_I^2 \ra_\tau$ as a function of
$\tilde{I} = I/I_c$ for both arrays of Fig.\ \ref{fig:10x_IV}, under
conditions such that all rows are active.  This plot is obtained by
following the decreasing current branch.  Surprisingly, when the
$10\times 10$ array (with $\tilde{g}_{(10\times 10)} = 0.1
~\tilde{g}_{(10\times 1)}$) locks on to the SIRS, $\tilde{E}$ jumps to
a value which is approximately two orders of magnitude {\em larger}
than that of the corresponding jump in the $10\times 1$ array, even
though the parameter $N_y\tg_x$ is the same for both arrays.  We
believe that the difference is due simply to the greater number of
junctions which are driving the cavity in the 2D case.  Even though
the {\em width} of the steps is controlled primarily by the parameter
$N_y\tg_x$, the energy in the cavity is determined by the {\em square}
of the number of radiating junctions.  This square is {\em 100 times
larger} for the 2D array than for the 1D array.

\section{Summary}

In this paper, we have derived equations of motions for a 2D array of
underdamped Josephson junctions in a single-mode resonant cavity,
starting from a suitable model Hamiltonian and including the effects
of both a current drive and resistive dissipation.  In the limit of
zero junction-cavity coupling, these equations of motion correctly
reduce to those describing a 2D array of resistively and capacitively
shunted Josephson junctions.

As in our previous 1D model, the present equations of motion lead to a
transition from incoherence to coherence, as a function of the number
of active rows $N_a$.  This transition again results from the
effectively mean-field-like nature of the interaction between the
junctions and the cavity.  Specifically, because each junction is, in
effect, coupled to every other active junction via the cavity, the
strength of the effective coupling is proportional to the number of
active junctions.  Thus, for any $\tilde{g}_x$, no matter how small, a
transition to coherence is to be expected for sufficiently large
number of active rows $N_a$.  We also found a striking effect of
polarization: the transition to coherence occurs only when the cavity
mode is polarized so that its electric field has a component parallel
to the direction of current flow.

Next, we briefly compare our numerical results to the behavior seen in
experiments.\cite{barbara,vasilic02} Our calculations show the
following features seen in experiments: (i) self-induced resonant
steps (SIRS's) in the IV characteristics; (ii) a transition from
incoherence to coherence above a threshold number of active junctions;
and (iii) a total energy in the cavity which varies quadratically with
the number of active junctions when those junctions are locked onto
SIRS's.  There may, however, be some differences as well.  In
particular, our transition to a quadratic behavior occurs when the
active junctions are locked onto SIRS's.  In possible contrast to our
results, in some experimental arrays,\cite{vasilic02} it has been
reported that even {\em below} the ``coherence threshold,'' individual
rows of junctions are locked onto SIRS's, but these SIRS's are not
coherent with one another, and hence, do not radiate an amount of
power into the cavity proportional to the square of the number of
junctions on the SIRS's.  Thus far, in our calculations, we have found
that when $N_a$ junctions are locked onto the steps, the energy in the
cavity is quadratic in $N_a$.  The threshold, in our calculations,
occurs when all the active junctions lock onto SIRS's, {\em not} when
active junctions which are already locked onto SIRS's become coherent
with one another.

For some choices of the parameters $\tg_x$, $\tilde{\Omega}$,
$\Delta$, $\beta$, and $\tilde{I}$, we find dynamical states such that
all active rows lock onto SIRS's while $\langle r \rangle_\tau < 1$.
In such states, the Kuramoto order parameter for the {\em individual
rows} is still $\langle r \rangle_\tau \sim 1$, implying that the rows
are not perfectly phase-locked to each other.  An example of such a
state is shown in Fig.\ \ref{fig:kuram_small}.  In such states, our
calculated energy $\tilde{E}$ in the cavity appears to vary smoothly
with $N_a$, and exhibits no threshold behavior, in contrast to what we
find at other applied currents [cf.\ Fig.\ \ref{fig:40x_thresh}].
This behavior appears to differ from what was reported experimentally
in a recent paper;\cite{vasilic02} the reasons for the difference are
not clear to us.

In summary, we have extended our previous theory of Josephson junction
arrays coupled to a resonant cavity to the case of 2D arrays.  The 2D
theory bears many similarities to the 1D case, and makes clear why the
1D model works so well.  These similarities arise because, in a square
array, the coupling to the cavity takes place only through those
junctions which are parallel to the applied current.  Again as in 1D,
our model leads to clearly defined SIRS's whose voltages are
proportional to the resonant frequency of the cavity.  Another
similarity is that, in 2D as in 1D, when a fixed number of rows are
biased on a SIRS, the cavity energy is linear in the input power.

We also find some results which are specific to 2D.  For example,
whenever one junction in a given row is biased on a SIRS, all the
junctions in that row phase-lock onto that same SIRS.  In addition,
the time-averaged energy contained in the resonant cavity is quadratic
in the number of active rows, but, when the array is biased on a SIRS,
is much {\em larger} in the 2D array than in the 1D array, for the
same value of the coupling parameter $\tg_x N_y$.

When the cavity mode is polarized perpendicular to the direction of
current drive, we find that the cavity does not affect the array IV
characteristics.  Our equations suggest that this non-effect might
change if the array were frustrated, e. g., by an external magnetic
field normal to the array.  Such frustration would cause junctions in
the $x$ and $y$ direction to be coupled.  It would be of great
interest if this speculation could be tested experimentally.

\begin{acknowledgments}
This work has been supported by the National Science Foundation,
through grant DMR01-04987, and in part by the U.S.-Israel Binational
Science Foundation.  Some of the calculations were carried out using
the facilities of the Ohio Supercomputer Center, with the help of a
grant of time.  We are very grateful for valuable conversations with
Profs. T. R. Lemberger and C. J. Lobb.
\end{acknowledgments}


\begin{thebibliography}{100}
\bibitem{note1} Note that one always has the relation $\tg_{ij} =
-\tg_{ji}$.
\end{thebibliography}
\end{document}